\documentclass[runningheads,showdefinitionlinks]{llncs}

\AddToHook{cmd/ttfamily/after}{\frenchspacing}

\usepackage[all]{nowidow}

\usepackage{ifthen}
\usepackage{siunitx}
\usepackage[T1]{fontenc}
\usepackage{graphicx}
\usepackage{hyperref}
\usepackage{xcolor}
\usepackage[numbers]{natbib}
\edef\citet{\noexpand\leavevmode
            \noexpand\protect
            \expandafter\noexpand\csname citet \endcsname}
\edef\citep{\noexpand\leavevmode
            \noexpand\protect
            \expandafter\noexpand\csname citep \endcsname}
\edef\cite{\noexpand\leavevmode
            \noexpand\protect
            \expandafter\noexpand\csname cite \endcsname}
\usepackage{listings}

\usepackage{xintbinhex}
\usepackage{xintexpr}

\def\linestep{4}
\def\makepclinenumbers{%
    \def\thelstnumber{%
        \ifnum\value{lstnumber}=0%
            i%
        \else%
            i+\xintDecToHex{\the\numexpr "\linestep * \the\value{lstnumber}\relax}%
        \fi%
}}

\newcommand*\makeaddresslinenumbers[1]{%
    \def\tmp{#1}%
    \def\thelstnumber{%
        0x\xintDecToHex{\the\numexpr \tmp + "\linestep * \the\value{lstnumber}\relax}%
}}

\def\makeplainlinenumbers{%
    \def\thelstnumber{%
    {\the\numexpr1+\the\value{lstnumber}}%
}}

\lstdefinelanguage{myasm}[x86masm]{Assembler}{
    morekeywords={define,cbz,ldr,bne,cbnz},
    deletekeywords=[2]{loop},
}
\lstdefinestyle{customasm}{
    belowcaptionskip=1\baselineskip,
    frame=single, 
    frameround=tttt,
    xleftmargin=\parindent,
    language=myasm,
    basicstyle=\scriptsize\ttfamily,
    commentstyle=\itshape\color{seabornGreen},
    keywordstyle=\color{seabornBlue},
    identifierstyle=\color{seabornRed},
    tabsize=2,
    numbers=left,
    numbersep=8pt,
    stepnumber=1,
    numberstyle=\ttfamily\tiny\color{gray}, 
    columns=fixed,
    firstnumber=0,
}

\usepackage[utf8]{inputenc}
\usepackage{tikz}
\usetikzlibrary{arrows,automata,positioning,calc}
\usepackage{wrapfig}
\usepackage{amsmath}
\usepackage{amssymb}
\usepackage{stmaryrd}
\usepackage{xargs}
\usepackage{xstring}
\usepackage{mathtools}
\usepackage{suffix} 
\usepackage{mathpartir}
\usepackage[firstpage]{draftwatermark} 
\usepackage{enumitem}
\setlist{nosep}

\newcommand*\anonymlink[2]{#2} 

\newcommand*\bignumreporawlink{https://github.com/\bignumuserraw/\bignumreporaw}
\newcommandx\definebignumlink[4][1=\bignumreporawlink]{\def\tmplink{#1/blob/\bignumcommithash/#2\#L#3-L#4}}

\newcommandx\makebignumlink[3]{\bignumreporawlink/blob/\bignumcommithash/#1\#L#2-L#3}

\newcommand*\bignum{\bignumreporaw} 
\newcommand*\bignumurl{\url{\bignumreporawlink}}

\NewDocumentEnvironment{hyperdefinition}{O{} m}
  {%
    \refstepcounter{definition}%
    \vspace{\topsep}%
    \par\noindent%
    \textbf{\href{#2}{Definition~\thedefinition\IfValueT{#1}{~(#1)}}.}\itshape%
  }
  {
    \par\vspace{\topsep}
  }

\definecolor{seabornBlue}{RGB}{76, 114, 176}
\definecolor{seabornOrange}{RGB}{221, 132, 82}
\definecolor{seabornGreen}{RGB}{85, 168, 104}
\definecolor{seabornRed}{RGB}{196, 78, 82}
\definecolor{seabornPurple}{RGB}{129, 114, 179}
\definecolor{seabornBrown}{RGB}{140, 86, 75}
\definecolor{seabornPink}{RGB}{144, 103, 167}
\definecolor{seabornGray}{RGB}{207, 207, 207}
\definecolor{seabornYellow}{RGB}{242, 192, 64}

\newcommand*\Section[1]{Section~\ref{sec:#1}}
\newcommand\Figure[2][]{Figure~\ref{fig:#2}\ifthenelse{\equal{#1}{}}{}{~(#1)}}

\newcommand*\Example[1]{Example~\ref{ex:#1}}

\newcommand*\Line[1]{Line~\ref{line:#1}}
\newcommand*\Appendix[1]{Appendix~\ref{app:#1}}
\newcommand*\Theorem[1]{Theorem~\ref{thm:#1}}
\newcommand*\Lemma[1]{Lemma~\ref{lemma:#1}}
\newcommand*\Definition[1]{Definition~\ref{def:#1}}

\newcommand*\arm{{\normalfont\textsc{arm}}}
\newcommand*\x[2]{{\normalfont\textsc{x86}}} 

\newcommand*\nat[1][]{\mathbb{N}_{#1}}

\newcommand*\true{\textsc{true}}
\newcommand*\false{\textsc{false}}

\newcommand*\byte{\texttt{byte}}

\renewcommand*\int[1][]{\texttt{int}{#1}}
\newcommand*\events{\mathbb{E}}
\newcommand*\load{\texttt{load}}
\newcommand*\store{\texttt{store}}
\newcommand*\branch{\texttt{branch}}

\newcommandx\elimset[3][1=s_0',2=0,3={Q}]{\dot{#3}_{\pi_{#2} = {#1}}}
\DeclareMathOperator*\iffdef{\overset{\text{\scalebox{.9}{def}}}{\iff}}
\DeclareMathOperator*\defeq{\overset{\text{\scalebox{.6}{\textup{def}}}}{=}}
\newcommand*\setdef[2]{\left\{#1 \mid #2\right\}}
\newcommand*\listdef[2]{\left[#1 \mid #2\right]}
\WithSuffix\newcommand\setdef*[2]{\left\{ #1 \left| \; \begin{array}[l]{@{}l@{}} #2 \end{array}\!\right.\right\}}
\WithSuffix\newcommand\listdef*[2]{\left[ \left. \begin{array}[l]{@{}l@{}} #1 \end{array}\;\right| #2 \right]}
\newcommand*\concat{+\!\!\!+}

\newcommand*\logic{\mathcal{L}_{2}}
\newcommand*\nonrellogic{\mathcal{L}_{1}}

\newcommand*\state{\Sigma}

\newcommand*\armlabels{\labels_{\arm}}

\newcommand*\eventlabels{\labels_{\normalfont \text{e}}}
\newcommand*\stepsymbol{\tau}
\newcommand*\stepssymbol{\stepsymbol^n}
\newcommand*\stepprodsymbol{\stepsymbol^2}
\newcommand*\steplocksymbol{\stepsymbol^{\text{lock}}}

\newcommand*\stepsymbolx{\stepsymbol_{\x86}}
\DeclareMathOperator*\step{\xrightarrow{\stepsymbol}}
\DeclareMathOperator*\xstep{\xrightarrow{\stepsymbolx}}

\DeclareMathOperator*\stepprod{\xrightarrow{\stepprodsymbol}}
\DeclareMathOperator*\steplock{\xrightarrow{\steplocksymbol}}
\DeclareMathOperator*\steps{\xrightarrow{\stepsymbol^n}}

\DeclareMathOperator*\stepsprime{\xrightarrow{\stepsymbol^{l}}}

\newcommand*\instr{\normalfont \texttt{instr}}
\newcommand*\pc{\normalfont \texttt{pc}}
\newcommand*\rip{\normalfont \texttt{rip}}
\newcommand*\registers{\normalfont \texttt{regs}}

\newcommand*\flags{\normalfont \texttt{flags}}

\newcommand*\memory{\normalfont \texttt{memory}}
\newcommand*\program{\normalfont \texttt{prog}}
\newcommandx\alignedto[4][1=s,2={i_0},3=C,4=\stepsymbol]{\normalfont \anonymlink{alignedto}{\textsc{align}^{#4}}(#1, #2, #3)}
\newcommandx\terminatedto[3][1=s,2=i,3=\stepsymbol]{\normalfont \anonymlink{terminatedto}{\textsc{end}^{#3}}(#1, #2)}
\newcommand*\maychange{\normalfont \anonymlink{maychange}{\textsc{mayChange}}}
\newcommandx\decode[3][1=s,2={i_0},3=\stepsymbol]{\normalfont \anonymlink{decode}{\textsc{decode}^{#3}}(#1, #2)}

\newcommand*\length[1][\stepsymbol]{\normalfont \anonymlink{length}{\textsc{length}^{#1}}}

\newcommand*\eventstate{\state_{\normalfont \text{e}}}
\newcommand*\labels{\mathbb{L}}
\newcommand*\publicname{\normalfont \text{pub}}
\newcommand*\public{\labels_{\publicname}}
\DeclareMathOperator*\equivpublic{{\simeq_{\publicname}}}
\newcommand*\privatename{\normalfont \text{pri}}
\newcommand*\private{\labels_{\privatename}}
\DeclareMathOperator*\equivprivate{{\simeq_{\privatename}}}
\newcommand*\stepsymbolevents{\stepsymbol^{\normalfont \text{e}}}
\DeclareMathOperator*\stepevents{\xrightarrow{\stepsymbolevents}}

\newcommand*\eventuallyname{\normalfont \texttt{eventually}}
\newcommandx\eventually[2][1={},2=Q]{\eventuallyname^{#1}\!\left({#2}\right)}
\newcommand*\eventuallynname{\normalfont \texttt{eventually{\color{seabornRed}n}}}
\newcommandx\eventuallyn[3][1={},2={n},3=Q]{\eventuallynname^{#1}_{#2}\!\left({#3}\right)}

\newcommand*\ensuresname{\normalfont \texttt{ensures}}
\newcommand*\ensurestwoname{\normalfont \texttt{ensures}{\color{seabornRed}2}}
\newcommand*\ensuresnname{\normalfont \texttt{ensures{\color{seabornRed}n}}}
\newcommandx\ensures[4][1={},2=P,3=Q,4=F]{\ensuresname^{#1}\!\left(#2, #3, #4\right)}
\WithSuffix\newcommand\ensures*[4]{\ensuresname^{#1}\!\left(\begin{array}[l]{@{}l@{}} #2,\\ #3,\\ #4 \end{array}\!\right)}
\newcommandx\ensurestwo[6][1={},2=\stepcalc0,3=\stepcalc1,4=P^\times,5=Q^\times,6=F^\times]{\ensurestwoname^{#1}_{#2,#3}{\!\left(#4, #5, #6\right)}}
\WithSuffix\newcommand\ensurestwo*[6]{\ensurestwoname^{#1}_{#2,#3}{\!\left(\begin{array}[l]{@{}l@{}} #4,\\ #5,\\ #6 \end{array}\!\right)}}
\newcommandx\ensuresn[5][1={},2=\stepcalc{},3=P,4=Q,5=F]{\ensuresnname^{#1}_{#2}{\!\left(#3, #4, #5\right)}}
\WithSuffix\newcommand\ensuresn*[5]{\ensuresnname^{#1}_{#2}{\!\left(\begin{array}[l]{@{}l@{}} #3,\\ #4,\\ #5 \end{array}\!\right)}}
\newcommand*\stepcalc[1]{f\!n_{#1}}

\newcommand*\hybridname{{\color{seabornRed}h}\ensurestwoname}
\newcommandx\hybrid[9][1={},2=\stepcalc0,3=\stepcalc1,4=P^\times,5=Q^\times,6=F^\times,7=P,8=Q,9=F]{\hybridname^{#1}_{#2,#3}{\!\left(#4, #5, #6 \mid #7, #8, #9\right)}}
\WithSuffix\newcommand\hybrid*[9]{\hybridname^{#1}_{#2,#3}{\!\left(\begin{array}[l]{@{}l@{}} #4,\\ #5,\\ #6 \end{array} \left| \begin{array}[l]{@{}l@{}} #7,\\ #8,\\ #9 \end{array}\right.\!\right)}}

\DeclareMathOperator*\equivin{{\simeq_{\text{in}}}}
\DeclareMathOperator*\equivout{{\simeq_{\text{out}}}}
\newcommand*\largestprefix{\textsc{largestPrefix}}
\newcommand*\eventuallynatpcname{\normalfont \texttt{eventually{\color{seabornRed}n}at{\color{seabornRed}pc}}}
\newcommandx\eventuallynatpc[4][1={n},2={x_0},3={x_\omega},4=P]{\eventuallynatpcname_{#1}^{#2,#3}\!\left({#4}\right)}
\WithSuffix\newcommand\eventuallynatpc*[4]{\eventuallynatpcname^_{#1}^{#2,#3}\!\left(\begin{array}[l]{@{}l@{}} #4 \end{array}\!\right)}
\newcommand*\equivensuresname{\normalfont \texttt{{\color{seabornRed} eq}\ensuresname}}
\newcommandx\equivensures[9][1={},2=C_0,3=C_1,4=\equivin,5=\equivout,6=\pc_0,7=\pc_0',8=\pc_1,9=\pc_1']{\equivensuresname^{#1}_{#6,#7,#8,#9}{\!\left(#2, #3, #4, #5\!\right)}}

\newcommand*\proga{\texttt{compare}}
\newcommand*\progb{\texttt{cst-compare}}

\let\oldtexttt\texttt
\renewcommand\texttt[1]{%
\ensuremath{\oldtexttt{#1}}%
}

\begin{document}
\title{Relational Hoare Logic \\ for Realistically Modelled Machine Code}

\def\companyraw{Amazon}
\def\bignumuserraw{awslabs}
\def\bignumreporaw{s2n-bignum}
\def\bignumcommithash{c747b1b66801e3975a8da502e18962838d3be945}

\author{Denis Mazzucato\thanks{Denis Mazzucato and Abdalrhman Mohamed contributed equally to this work.}\inst{1}\orcidID{0000-0002-3613-2035}
\and
Abdalrhman Mohamed{\protect\footnotemark[1]}\inst{2}\orcidID{0000-0003-1414-7073}
\and
Juneyoung Lee\thanks{Juneyoung Lee is the corresponding author: \email{lebjuney@amazon.com}}\inst{3}\orcidID{0000-0002-8152-9330}
\and
Clark Barrett\inst{2}\orcidID{0000-0002-9522-3084}
\and
Jim Grundy\inst{3}\orcidID{0009-0006-5072-9520}
\and
John Harrison\inst{3}
\and
Corina S. P\u{a}s\u{a}reanu\inst{1}\orcidID{0000-0002-5579-6961}}
\authorrunning{D. Mazzucato et al.}
\institute{
  Carnegie Mellon University \\
  \email{\{dmazzuca,pcorina\}@andrew.cmu.edu}
  \and
  Stanford University \\
  \email{\{abdal,barrettc\}@stanford.edu}
  \and
  Amazon Web Services \\
  \email{\{lebjuney,jmgruj,jargh\}@amazon.com}
}


\SetWatermarkAngle{0}
\SetWatermarkText{\raisebox{9.6cm}{ 
\hspace{-.24cm}
\href{https://doi.org/10.5281/zenodo.15309209}{\includegraphics{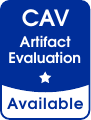}}
\hspace{8.42cm}
\includegraphics{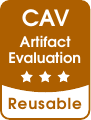}
}}

\maketitle

\begin{abstract}
  %
  Many security- and performance-critical domains, such as cryptography, rely on low-level verification to minimize the trusted computing surface and allow code to be written directly in assembly. However, verifying assembly code against a realistic machine model is a challenging task. Furthermore, certain security properties---such as constant-time behavior---require relational reasoning that goes beyond traditional correctness by linking multiple execution traces within a single specification. Yet, relational verification has been extensively explored at a higher level of abstraction. In this work, we introduce a Hoare-style logic that provides low-level, expressive relational verification.
  We demonstrate our approach on the \bignum{} library, proving both constant-time discipline and equivalence between optimized and verification-friendly routines.
  Formalized in HOL Light, our results confirm the real-world applicability of relational verification in large assembly codebases.
  \end{abstract}

\section{Introduction}
\label{sec:introduction}

Verification of low-level program properties is paramount for security-critical systems. This applies to microkernels \cite{klein2009}, where processors and other hardware may have effects that are not captured by high-level abstractions, as well as cryptographic libraries \cite{bond2017,barthe2018}, which aim to minimize the trusted computing base and build-toolchain dependencies. Additionally, performance-critical code is often written directly in assembly to maximize performance.

Many challenges arise when verifying low-level code, as programs execute on machines with finite memory, bounded integers, unstructured control flow, and memory space that is shared between data and code.
In contrast, high-level verification uses abstractions that simplify reasoning and hide hardware-specific details.
For instance, low-level verification must consider that primitives, when storing data, need to have free memory space.
Furthermore, the verification process becomes much harder when dealing with \emph{relational properties} \citep{clarkson2008}.
Relational properties link multiple execution traces together within a single property specification.
They are necessary for critical applications, such as proving that a cryptographic routine runs in constant time with respect to secret data, or that two versions of a program are functionally equivalent.

In this work, we target the \bignum{} library,\footnote{\bignumurl} a cryptographic library written in assembly for \arm{} and \x86{} architectures.  It includes mathematical operations on large integers, such as modular multiplication, as well as more cryptographic-oriented operations, such as elliptic curve operations.
As part of the AWS's TLS/SSL implementation, these arithmetic routines are both performance- and security-critical.
The library features both highly optimized routines that are hard to verify as well as verification-friendly variants that are easier to verify but slower in practice.
By verifying the latter and proving that they are functionally equivalent to the former, we can ensure that the high-performance versions do not compromise correctness.
We also aim to ensure that the high-performance routines execute in constant time, a property necessary to prevent timing side-channel attacks, which could compromise sensitive data.

A number of works previously studied Hoare-style logics for realistically modelled machine code \citep{wang1976a,protzenko2020,myreen2007b,lundberg2020,bartels2014,bosamiya2020}.
Hoare-style reasoning has also been pervasively studied for relational properties \citep{benton2004,rinard1999,blatter2022}.
However, relational verification for low-level code remains underexplored, especially for machine code with realistic features such as finite memory and unstructured control flow.
Ideally, a binary verification toolkit would use a robust Hoare-style logic that supports realistic machine code, can express relational properties, provides sound and complete proof rules, and retains key properties that users naturally expect. These include, for instance, commutativity, as well as the ability to weaken and strengthen pre- and postconditions and to unify contracts across different contexts.
Such \emph{natural properties} enable modular reasoning, support multiple proof strategies, and make the framework practical for real-world applications.
To the best of our knowledge, no existing work has presented a relational Hoare logic for realistically modelled machine code that satisfies all these properties.

In this paper, we fill that gap by introducing a novel Hoare-style logic for low-level, relational verification. Our framework, fully formalized in the HOL Light theorem prover \cite{harrison2009,harrison2011}, offers proof rules designed to meet users' natural expectations.
We demonstrate our approach via two major case studies: in the first, we show how our framework can be used to verify constant-time behavior of various routines; in the second, we show it can be used to prove the functional equivalence of two different implementations of the same routine (e.g., one optimized for speed and the other optimized for verifiability).
These case studies are conducted on the \bignum{} cryptographic library. Results show that our logic scales to large assembly programs and yields practical value.

While our primary application is the \bignum{} library, the generality of our relational Hoare logic extends beyond cryptographic code. It supports low-level features including indirect branches and self-modifying code, even though cryptographic libraries such as \bignum{} may not employ these features.

We summarize our contributions as follows:%
\begin{enumerate}[label=\roman*)]%
  \item A novel relational Hoare logic tailored to realistically modelled machine code, formalized in HOL Light.
  \item A first case study on constant-time behavior of cryptographic routines  in the \bignum{} library, including the copy and modular inversion routines.
  \item A second case study involving equivalence proofs between optimized and verification-friendly implementations of \bignum{} routines.
\end{enumerate}

\section{Related Work}\label{sec:related-work}

\paragraph{Hoare-Style Reasoning for Realistically Modelled Machine Code.}
Verifying realistically modelled machine code is challenging due to unstructured control flow, which traditional Hoare logics \citep{hoare1969} struggle to handle.
While \citet{affeldt2012} uses Hoare logic to verify low-level arithmetic routines, their work is limited to assembly fragments with structured control flow.
Several approaches address unstructured control flow, such as the inductive assertion method \citep[Section 2]{naumann2020} used by \citet{barthe2005} and \citet{lehner2007} to generate verification conditions, and the program logic by \citet{tan2015} based on continuation-passing style reasoning \citep{benton2005}. However, these methods often fail to specify pre- and postconditions with shared continuation labels.

Other notable efforts include a logic for total correctness of communicating unstructured programs \citep{bartels2014,jahnig2015,jahnig2016}, formalized in Isabelle/HOL, and one for reasoning about MIPS assembly in Coq~\citep{marti2008}. However, these logics are compositional only for nonoverlapping fragments. Full compositionality is critical for modular reasoning, which, in turn, is needed for scalability.

\citet{wang1976a} proposes a logic for total correctness of unstructured programs with multi-exit postconditions, but it does not guarantee postconditions upon first encounter.
Unstructured programs may, in fact, go through the last instruction, jump back, and then meet the postcondition later.
While this might seem misleading, we argue that functional specifications are generally confined to function boundaries, where the final instruction is a return statement.
This ensures the program cannot continue execution and revisit the postcondition later, effectively solving the issue of first-met postconditions.

\citet{myreen2007} introduce a logic for unstructured code applied in the verified CakeML compiler \citep{kumar2014}, leveraging decompilation into logic \citep{myreen2007b,myreen2007c,myreen2008,myreen2009,myreen2012}. Despite its major impact in verifying seL4 compilation \citep{sewell2013} and realistic executables~\citep{tan2015}, it lacks
a conjunction rule, a property that is naturally expected from a Hoare-style logic in order to unify contracts over different postconditions.
The lack of a conjunction rule significantly increases the proof burden.
\begin{wrapfigure}[5]{r}{0.25\textwidth}
  \centering
  \begin{minipage}{0.2\textwidth}
\begin{lstlisting}[
      style=customasm,
      morekeywords={loop},
      escapechar=|
      ]
  mov x0,xzr
loop: | \label{line:first:pre} |
  add x0,x0,#1
  cmp x0,#3 | \label{line:first:post} |
  bne loop
\end{lstlisting}
    \end{minipage}
\end{wrapfigure}
For instance, this assembly program 
increments \texttt{x0} by~1 until it reaches 3, then halts.
  Let $P = (\texttt{x0} = 1)$, $Q = (\texttt{x0} = 2)$, and $Q' = (\texttt{x0} = 3)$.
  The program satisfies $Q$ and $Q'$ separately, on line 3, after two and three iterations, respectively, but $Q$ and $Q'$ cannot possibly
  hold simultaneously.
In \Section{relational}, we show how to handle such cases.

\citet{ray2008} address conjunction rules and first-met postconditions by tracking execution steps, while \citet{lundberg2020} extend this to handle multi-exit locations, ensuring postconditions hold at the first encounter. However, their approaches assume deterministic semantics, incompatible with architectures like \x86{}. \citet{jensen2013} addresses this gap by using separation logic \citep{ohearn2001,reynolds2002} but only for a subset of \x86{} code.
Furthermore, EverCrypt \citep{protzenko2020} verifies cryptographic primitives using a Hoare-style logic through C and assembly code interoperability, but it does not support \arm{} architecture. 
Similar to our embedding of relational to unary Hoare triples, exploiting the event list, EverCrypt is able to prove constant-time.
Fiat-Crypto \citep{Erbsen2019} generates verified cryptographic code from high-level specifications with applications to big-number arithmetic, in scope similar to the \bignum{} library.
All these approaches 
lack a robust foundation for generic relational properties---not only the ones reducible to unary properties.

\paragraph{Relational Hoare Logics.}
Relational Hoare logics \citep{naumann2020} extend unary Hoare triples to reason about multiple execution traces. \citet{benton2004} gives a relational Hoare logic for two execution traces, later generalized by \citet{blatter2022} to any number of traces. While Benton also
covers relational properties of low-level unstructured code \citep{benton2005}, their logic relies on an idealized computational model. We propose a generic framework for manual proof of relational properties.

In credible compilation, \citet{rinard1999} developed relational logics for pointer allocation, and \citet{benton2018} proposes a sound-but-incomplete fully automatic tool for equivalence preservation of compiled programs with minor differences. They verify HHVM bytecode, which is not a high-level language but not as low as assembly; for instance, they do not handle physical registers. Instead, our approach sacrifices automation, gaining both soundness and completeness.

\citet{barthe2011} propose product program constructions for equivalence reasoning, later extended \citep{beringer2011,antonopoulos2017} to support equivalence across multiple programs.
\citet{kang2018} describe a relational logic for LLVM code which lacks support for indirect branches and self-modifying code. \citet{pit-claudel2022} further extend relational verification to low-level stack machines.
Our logic handles relational properties but does not trade off any low level features of assembly code.

\section{Running Example}

\begin{figure}[t]
  \centering
  \makeplainlinenumbers
  \begin{minipage}[t]{0.44\textwidth}
\begin{lstlisting}[
      style=customasm,
      morekeywords={loop,eq,neq},
      caption={\proga},
      escapechar=|
      ]
  cbz n, eq
loop:
  sub n, n, #1
  ldr kn, [k, n, lsl #3]
  ldr xn, [x, n, lsl #3]
  cmp kn, xn
  bne neq
  cbnz n, loop
eq:
  mov res, #1
  ret
neq:
  mov res, xzr
  ret
\end{lstlisting}
    \end{minipage}
    \hfill
    \begin{minipage}[t]{0.44\textwidth}
\begin{lstlisting}[
        style=customasm,
        morekeywords={loop,end},
        caption={\progb},
        escapechar=|
        ]
  mov diff, xzr
  cbz n, end |\label{line:compare:cbz}|
loop:
  sub n, n, #1
  ldr kn, [k, n, lsl #3] |\label{line:compare:k}|
  ldr xn, [x, n, lsl #3] |\label{line:compare:x}|
  eor temp, kn, xn
  orr diff, diff, temp
  cbnz n, loop             |\label{line:compare:loop}|
end:
  cmp diff, xzr
  cset res, eq
  ret
\end{lstlisting}
      \end{minipage}
      \caption{Two programs to perform byte-by-byte comparison of buffers.
        The program \texttt{compare} (left) is not constant-time, while the program \texttt{compare-constant} (right) is.
      }
        \label{fig:running-example}
  \end{figure}

We use the program $\proga{}$, shown in \Figure[left]{running-example}, as a running example for the rest of the paper. It compares byte-by-byte the contents of
a key buffer $\vec{k}$ and a data buffer $\vec{x}$ of length $n$.
It takes as input the buffer length $n$ and the memory addresses of the buffers $k$ and $x$, provided via the registers $\texttt{n}$, $\texttt{k}$, and $\texttt{x}$, respectively. Temporary values are stored in the registers $\texttt{kn}$, $\texttt{xn}$, $\texttt{diff}$, and $\texttt{temp}$, while the result is stored in $\texttt{res}$.
The private data is the content of the key buffer.
The program $\proga{}$ iterates backwards, comparing corresponding elements from both buffers. If a mismatch is detected, the program jumps to the label $\texttt{neq}$ and sets $\texttt{res}$ to $0$. Otherwise, if all elements match, it reaches the label $\texttt{eq}$ and sets $\texttt{res}$ to $1$. This behavior results in variable execution time depending on the buffer contents. An attacker can exploit this timing variation to deduce the position of mismatches and reconstruct the secret buffer $\vec{k}$ in linear time.

To address this issue, program $\progb{}$ in \Figure[right]{running-example} implements a constant-time comparison. It always iterates over the entire buffer length, regardless of mismatches, accumulating possible differences in $\texttt{diff}$. The program's execution time is constant for any buffer content, ensuring no timing leaks and preventing attackers from inferring secret information. Furthermore, the two programs are functionally equivalent.

\section{Unary Hoare Logic $\nonrellogic$}\label{sec:unary}

This section provides background on an unary Hoare logic
used in the verification framework described in
\citep{myreen2007,lundberg2020}.  We refer to this logic as $\nonrellogic$. The definitions and theorems of $\nonrellogic$ have been fully mechanized in HOL Light by previous researchers.

\paragraph{States.}
Let $\state$ denote a set of machine states, represented as the set of functions mapping observable resources $\labels$ (e.g., memory, registers, program counter) to their values. For example, in the \arm{} architecture, the resources $\armlabels$ include a 64-bit program counter $\pc$, 32 general-purpose registers $\registers_i$, flags $\flags_k$, and memory $\memory_h$, indexed accordingly by $i$, $k$, and $h$.
Similarly, the \x86{} architecture has resources like an instruction pointer ($\rip$) and extended flags. To generalize across different architectures, the label $\instr$ refers to the address of the next instruction, where $\instr = \pc$ for \arm{} and $\instr = \rip$ for \x86{}.
%
We use $s(l)$ for the value of resource $l\in\labels$ in the state $s\in\state$ and $s[l \mapsto v]$ for the updated state.
Resource values depend on the architecture; e.g., for \arm{}, $s(\pc) \in \int[64]$, $s(\registers_i) \in \int[64]$, and $s(\memory_h) \in \byte$, where $\byte \defeq \{0,1\}^8$ and $\int[n] \defeq \{0,1\}^n$.

\paragraph{Properties.} A property $P$ is a subset of machine states $\state$. A state $s$ satisfies the property $P \subseteq \state$ if $s \in P$. The execution of a single instruction is modelled as the small-step operational semantics $\stepsymbol \subseteq \state \times \state$, where $s \step s'$ describes the fetch-decode-execute cycle, updating state $s$ to $s'$ and advancing $\instr$. The composition of two relations $\stepsymbol_1, \stepsymbol_2$ is defined as $\stepsymbol_1 \circ \stepsymbol_2 \defeq \{(s, s'') \mid \exists s'.~ s \xrightarrow{\stepsymbol_1} s' \xrightarrow{\stepsymbol_2} s'' \}$.
The $n$-th composition of $\stepsymbol$ is $\stepssymbol$.
The decoding function $\decode[s][i]$ maps bytes in the memory at address $i$
either to an instruction or to $\bot$ if undecodable. \arm{} instructions have a length of 4 bytes and are 4-byte aligned. \x86{} instructions have variable lengths.
We write $\length(C)$ for the number of bytes that the program $C$ occupies in the memory without padding for alignment.
Execution halts when the first undecodable instruction is encountered, denoted by $
\terminatedto[s][i] \iffdef s(\instr) = i \land \decode[s][i] = \bot$.

We use $\alignedto$ in this paper to denote that the program $C$ is stored in memory starting from the address $i_0$ where $s(\instr)= i_0$ and $i_0$ satisfies the alignment constraint of a program if the architecture is \arm{}.
The predicate $\alignedto$ may appear as a conjunctive clause in $P$ to describe the program of interest. The notation $\program(P)$ refers to the program $C$ constrained by $P$.

\paragraph{The $\eventuallyname$ Property.}
Assume that a machine state $s \in \state$ satisfies a precondition.
A postcondition $Q \subseteq \state$ must eventually hold after a finite number of steps from $s$.
To represent such $s$, $\eventually[\stepsymbol][Q]$ defines the set of states from which $Q$ eventually
holds along every possible path through $\stepsymbol$.

%
%

\begin{definition}[Eventually][\makebignumlink{common/relational.ml}{994}{998}]\label{def:eventually}
  Given an operational semantics $\stepsymbol\subseteq \state \times \state$ and a property $Q\subseteq \state$, the property $\eventually[\stepsymbol]\subseteq \state$ is defined inductively as:
    \begin{mathpar}
        \inferrule{s \in Q}{s\in\eventually[\stepsymbol]}
        \and
        \inferrule{
          \exists s'.~ s \step s'
        \\ \forall s'. \, s \step s' \implies s' \in \eventually[\stepsymbol][Q] }{s\in\eventually[\stepsymbol]}
    \end{mathpar}
\end{definition}


The second inference rule expands $\eventuallyname$\footnote{We omit the $\stepsymbol$ symbol when the operational semantics is clear from the context.} if every next state $s'$ is in $\eventually[\stepsymbol]$.
This notion of $\eventuallyname$ is essential for reasoning about nondeterministic operational semantics, such as in \x86, where certain instructions exhibit nondeterministic behavior.
For instance, the \texttt{mul}
instruction\footnote{\url{https://www.felixcloutier.com/x86/mul\#flags-affected}}
nondeterministically sets the \texttt{SF} flag to either 0 or 1.
A simplified small-step semantics for \texttt{mul} is as follows:
\begin{align*}
  \resizebox{\linewidth}{!}{$
  \begin{aligned}
\inferrule*[right=mul]{
  s(\instr) = i
  \quad
  \decode[s][i][\stepsymbolx] = \texttt{mul}~r
  \quad
  r \in \int[16]
  \quad
    s(r) = x
    \quad
    {s\!f} \in \{0, 1\}
}{
    s \xstep s\left[
        \texttt{EAX} \mapsto s(\texttt{AX}) \cdot x,~
        \texttt{SF} \mapsto {s\!f},~
        \instr \mapsto i + \length(\texttt{mul}~r)
    \right]
}
\end{aligned}
  $}
\end{align*}

\begin{example}\label{ex:non-det}
  Consider a program $C$ consisting of the instructions \texttt{mul\;ax}, \texttt{sets\;dl}, and \texttt{imul\;edx,\;eax}. The program starts at instruction register $i_0$, where \texttt{AX} (the
  least significant 16 bits of \texttt{EAX}) is multiplied by itself, setting \texttt{SF} to 0 or 1. In the second instruction, the least significant byte of \texttt{EDX}, referred to as \texttt{DL}, is set to \texttt{SF}.
  After then, the values of \texttt{EAX} and \texttt{EDX} are multiplied, and the truncated result up to 32 bits is stored in \texttt{EDX}.
  The program terminates at $i_0+9$ because each of the \x86 instructions is 3 bytes, and \texttt{EDX} is equal to either $0$ or $x^2$.
  The postcondition can be expressed as:
  $\setdef*{s'}{
  \terminatedto[s'][i_0 + 9][\stepsymbolx] \land (s'(\texttt{EDX}) = x^2 \lor s'(\texttt{EDX}) = 0)}$.
  It holds under a precondition requiring \texttt{AX} to initially be equal to $x$ and \texttt{EDX} to $0$.
\end{example}

\paragraph{Unary Hoare Triple.}


Reasoning about machine code differs from reasoning about high-level languages
in several ways.
First, the machine code is not represented as a syntactic program but instead as a set of instructions in the memory space.
Second, a machine code may modify anything during its execution, including itself and callee-save registers.
To denote unwanted modifications after the program execution, a \emph{frame} condition $F \subseteq \state \times \state$ bounds allowed changes of state components between the input and output states.
%
We explain the formal definition of the predicate $\ensuresname$ which is the Hoare triple in $\nonrellogic$. The notation used in this paper follows the convention you may find in the \bignum{} library.

\begin{definition}[Ensures][\makebignumlink{common/relational.ml}{1164}{1167}]\label{def:ensures}
    Given an operational semantics $\stepsymbol\subseteq \state \times \state$, a precondition $P\subseteq \state$, a postcondition $Q\subseteq \state$, and a frame condition $F\subseteq \state \times \state$, we define the predicate $\ensures[\stepsymbol]$ as follows:
    \begin{align*}
        &\ensures[\stepsymbol] \iffdef \\
        &\quad \forall s.~ s \in P \implies
            s \in \eventually[\stepsymbol][
                \setdef{s'}{s' \in Q \land (s, s') \in F}
            ]
    \end{align*}
\end{definition}

\begin{example}
Consider the program $C$ from \Example{non-det},
starting from the precondition $\setdef{s}{
    \alignedto[s][i_0][C][\stepsymbolx] \land s(\texttt{AX}) = x \land s(\texttt{EDX}) = 0
}$, ensuring that the memory is aligned with $C$ and $\texttt{AX}$ is equal to $x$.
By application of the operational semantics, we eventually satisfy the postcondition where $C$ terminates with \texttt{EDX} equal to 0 or $x^2$.
During execution, $C$ may modify \texttt{EAX}, \texttt{EDX}, and the sign flag \texttt{SF}.
We denote by $\maychange: \wp(\labels) \to \wp(\state \times \state)$ the resources that the program may modify.
Formally, $\hypertarget{maychange}\maychange(L) \defeq \setdef{(s, s')}{\forall l \in \labels.~ l \not\in L \implies s(l) = s'(l)}$.
    Thus, the frame condition can be written as
  \begin{math}
\maychange(\{\instr, \texttt{EAX}, \texttt{EDX}, \texttt{SF}\})
  \end{math}.
%
  The correctness of $C$ is captured by:
    \begin{align*}
        \ensures*{
          \stepsymbolx}{
          \setdef{s}{
            \alignedto[s][i_0][C][\stepsymbolx] \land s(\texttt{AX}) = x \land s(\texttt{EDX}) = 0
          }}{
            \setdef{s}{
              \terminatedto[s][i_0 + 12][\stepsymbolx] \land \left(s(\texttt{EDX}) = x^2 \lor s(\texttt{EDX}) = 0\right)
            }
          }{
            \maychange(\{\instr, \texttt{EAX}, \texttt{EDX}, \texttt{SF}\})
          }
    \end{align*}
\end{example}


Recall that the frame rule in separation logic \citep{ohearn2001,reynolds2002} states that if $\{P\}~C~\{Q\}$ holds, then for a disjoint memory region $R$, $\{P \ast R\}~C~\{Q \ast R\}$ also holds. Similarly, in our logic, if $R$ is invariant under $\maychange(L)$, i.e.,
$\forall s, s'.~ (s, s') \in \maychange(L)\! \implies\! (s \in R\! \iff\! s' \in R)$
then $\ensures[][P][Q][\maychange(L)]$ implies $\ensures[][P \cap R][Q \cap R][\maychange(L)]$. Therefore, $\nonrellogic$ supports modular verification while preserving the simplicity of first-order predicates, enabling efficient proof automation.

The logic $\nonrellogic$ is equipped with the usual derivation rules for reasoning about the program execution, cf. \Appendix{nonrellogic}.
The logic core and tactics are implemented in 10k lines of HOL Light \citep{harrison2009}.
It is currently used to verify functional safety properties of the \bignum{} library, comprising 615 arithmetic routines written in \arm{} and \x86{} assembly languages for P-256/384/521, x25519/ed25519 and RSA.
A total of 1013 functional properties have been verified, amounting to 860k lines of proofs.

\section{Program Logic \texorpdfstring{$\logic$}{L_2} for Relational Verification}
\label{sec:relational}

In this section, we first introduce a stronger variant of the $\eventuallyname$ predicate. Then, we present the relational logic $\logic$ as a natural extension of $\nonrellogic$.
We show how to prove a unary Hoare triple from a relational one and vice versa. This last step is essential in demonstrating the robustness of our logic and allows proofs to transition between $\nonrellogic$ and $\logic$.
We highlight the main extensions that allow us to prove relational properties and leave a discussion about the details of the challenges in \Appendix{challenges}.

\subsection{Unary Hoare Triples with Number of Steps}
Building on \citep{ray2008}, we propose a stronger $\eventuallyname$ operator that explicitly specifies the number of steps required to reach a given postcondition.


\begin{definition}[Stronger Eventually][\makebignumlink{common/relational_n.ml}{155}{159}]
  \label{def:eventually_n}
    Given an operational semantics $\stepsymbol\subseteq \state\times\state$ and a number of steps $n\in\nat$, for any postcondition $Q\subseteq\state$, we define:
    \begin{align*}
        &\eventuallyn[\stepsymbol] \defeq \setdef*{s\in\state}{
        \forall s'.~ s \steps s' \implies s' \in Q \land {} \\
            \forall s', l \in \nat.~ l < n \land s \stepsprime s'
            \implies \exists s''.~ s' \step s''}
    \end{align*}
\end{definition}

\noindent
That is, it defines the set of states such that for all states reachable in $n$ steps, the postcondition $Q$ must hold, and for all states reachable in less than $n$ steps, there must exist a successor state.

There are two merits in specifying the number of steps $n$.
First, it makes the conjunction rule sound.
In low-level languages, a program execution that failed to satisfy the postcondition at $\instr$ may continue as long as it encounters decodable instructions, and then branch back prior to $\instr$ and eventually satisfy the postcondition.
Therefore, writing multiple postconditions at $\instr$ that hold at different steps but not together would break the conjunction rule as shown in \Section{related-work}.
Explicitly stating the exact number of steps to arrive at the postcondition as an additional constraint resolves such problem.
Second, it retains soundness of the commutativity and composition of nested $\eventuallynname$ operators, which are similarly important for proving natural properties of relational Hoare triples.
When a low-level program exhibits nondeterministic behavior, each trace may meet the postcondition after different numbers of steps.\footnote{\scriptsize\url{\makebignumlink{common/relational2.ml}{86}{243}}}
The definition of $\eventuallynname$ is stronger than $\eventuallyname$, cf. \Definition{eventually}.
\begin{lemma}[][\makebignumlink{common/relational_n.ml}{268}{289}]\label{lemma:stronger_eventually}
  \begin{math}
    \forall Q\subseteq \state, n_{} \in \nat.~
    \eventuallyn[][n_{}] \subseteq \eventually[]
  \end{math}
\end{lemma}

The stronger eventually operator supports the following properties:
\begin{description}
  \item[\href{\makebignumlink{common/relational_n.ml}{171}{175}}{Conjunction}] As we require postconditions to hold after exactly $n$ steps, we can unify contracts stating different postconditions on the final states.
  \[
  \inferrule*[right=conj]{s \in \eventuallyn[][n] \\ s \in \eventuallyn[][n][Q']}{
      s \in \eventuallyn[][n][Q \cap Q']}
  \]
  \item[\href{\makebignumlink{common/relational2.ml}{249}{254}}{Commutativity}] Nested eventually operators commute, implying that the order of the two programs specified by the relational property will not matter. Whenever $Q^\times\subseteq \state\times\state$ is eventually satisfied in $n_0$ and $n_1$ steps, for the first and second components of $Q^\times$, the inverse $\setdef{(s_1, s_0)}{(s_0, s_1) \in Q^\times}$ is satisfied in $n_1$ and $n_0$ steps, respectively.
  \[
  \inferrule*[right=comm]{
    s_0 \in \eventuallyn[][n_0][{
      \setdef*{s_0'}{s_1 \in \eventuallyn[][n_1][
        \elimset^\times
      ]}
    }]
  }{
    s_1 \in \eventuallyn[][n_1][{
      \setdef*{s_1'}{s_0 \in \eventuallyn[][n_0][
        \ensuremath{\elimset[s_1'][1]^\times} 
      ]}
    }]
  }
  \]
Here, $\elimset[s_x][i]^\times\subseteq \state$ contains all states that satisfy $Q$ together with $s_x$ in the $i$-th component, i.e.,
$\elimset[s_x][i]^\times \defeq \setdef{
  \pi_{1-i}(s, s')
}{
   \pi_i(s, s') = s_x \land (s, s')\in Q^\times
}$.
The projection $\pi_i$ retrieves the $i$-th component of a pair of states (zero indexed).
Projections are lifted to sets of states by $\pi_i(Q^\times) \defeq \setdef{\pi_i(s, s')}{(s, s') \in Q^\times}$.

  \item[\href{\makebignumlink{common/relational2.ml}{256}{296}}{Composition}] Two fragments reaching $Q^\times$ and $R^\times$ in $n_0$, $n_1$ and $m_0$, $m_1$ steps, respectively, can be composed to reach $R^\times$ in $n_0 + m_0$ and $n_1 + m_1$ steps.
  \begin{align*}
    \resizebox{\linewidth}{!}{$
    \begin{aligned}
      \inferrule*[right=comp]{
        s_0\in\eventuallyn[][n_0][{
          \setdef*{s}{s_1\in\eventuallyn[][{n_1}][
            \ensuremath{\elimset[s][0]^\times}
        ]}
        }]
        \\
        \forall s_0', s_1'. (s_0', s_1')\in Q \!\implies\! s_1'\in\eventuallyn[][m_0][{\!
          \setdef*{\!s\!}{\!s_0'\in\eventuallyn[][{m_1}][
            \ensuremath{\elimset[s][0][R]^\times}
          ]\!}\!
        }]
      }{
        s_0\in\eventuallyn[][n_0 + m_0][{
          \setdef*{s}{s_1\in\eventuallyn[][{n_1 + m_1}][
            \ensuremath{\elimset[s][0][R]^\times}
          ]}
        }]
      }
    \end{aligned}
  $}
  \end{align*}

\end{description}

With the three properties of $\eventuallynname$ (cf. \textsc{conj}, \textsc{comm}, and \textsc{comp}), we can define a unary Hoare triple that maintains the properties that users would naturally expect from a Hoare logic.
To do so, we employ a step function $\stepcalc{}:\state\to\nat$ to make the number of steps dependent on a given state.

\begin{definition}[Stronger Ensures][\makebignumlink{common/relational_n.ml}{350}{353}]\label{def:ensures_n}
  Given an operational semantics $\stepsymbol \subseteq \state \times \state$, a precondition $P \subseteq \state$, a postcondition $Q \subseteq \state$, a frame condition $F \subseteq \state \times \state$,
  and a step function $\stepcalc{}:\state\to\nat$, a \emph{unary Hoare triple} is a statement of the form $\ensuresn$, where:
  \begin{align*}
    &\ensuresn \iffdef {} \\
    &\quad \forall s.~ s \in P \implies s \in  \eventuallyn[\stepsymbol][\stepcalc{}(s)][
      \setdef*{s'}{s' \in Q \land (s, s') \in F}
    ]
  \end{align*}
\end{definition}
Whenever precondition $P$ holds for state $s$, postcondition $Q$ holds for any state $s'$ that is related by $\stepcalc{}(s)$ steps of the execution of the program $\program(P)$, and $\program(P)$ modifies only the memory locations specified by the frame condition $F$.

As a consequence of \Lemma{stronger_eventually}, the unary Hoare triple $\ensuresn$ is stronger than $\ensuresname$, cf. \Definition{ensures}.

\begin{theorem}[][\makebignumlink{common/relational_n.ml}{363}{366}]\label{thm:ensuresn-ensures}
  \begin{math}
    \forall P, Q, F, \stepcalc{}.~
    \ensuresn \implies \ensures
  \end{math}
\end{theorem}

The other direction of the implication is not always true; in fact, it holds only for deterministic programs.
The reason is that the program may branch based on a nondeterministic choice, and the postcondition may hold in a different number of steps than the one specified in the Hoare triple.

\begin{theorem}[][\makebignumlink{common/relational_n.ml}{394}{399}]\label{thm:ensures-ensuresn}
  For any operational semantics $\stepsymbol$, precondition $P$, postcondition $Q$, frame condition $F$, if $\stepsymbol$ is deterministic, then:
  \begin{gather*}
    \ensures[\stepsymbol] \implies \exists \stepcalc{}.~  \ensuresn[\stepsymbol]
  \end{gather*}
\end{theorem}

\subsection{Relational Hoare Triples}

We now define the relational Hoare triple $\ensurestwo$ that allows us to reason about the behavior of two programs.
Whenever the precondition $P^\times\subseteq \state\times\state$ holds for a pair of states $(s_0, s_1)$, the postcondition $Q^\times\subseteq \state\times\state$ should eventually hold for any pair of states $(s_0', s_1')$ that are related by respectively $\stepcalc0$ and $\stepcalc 1$ steps of
the execution of the two programs $C_0$ and $C_1$.
As for the logic $\nonrellogic$, the two programs are not explicitly given but instead are constrained in the memory space by $P$, i.e., $C_0 = \program(\pi_0(P))$ and $C_1 = \program(\pi_1(P))$.
The frame condition $F^\times\subseteq (\state\times\state)\times(\state\times\state)$ specifies the memory locations that can be modified by the two programs.
Formally:

\begin{definition}[Relational Ensures][\makebignumlink{common/relational2.ml}{22}{28}]\label{def:ensures_1}
    Given operational semantics $\stepsymbol \subseteq \state \times \state$, a precondition $P^\times \subseteq \state \times \state$, a postcondition $Q^\times \subseteq \state \times \state$, and a frame condition $F^\times \subseteq (\state \times \state) \times (\state\times\state)$,
    two step functions $\stepcalc0, \stepcalc1 :\state\to\nat$,
    a \emph{relational Hoare triple} is a statement of the form $\ensurestwo$, where:
  \begin{align*}
    \resizebox{\linewidth}{!}{$
    \begin{aligned}
        &\ensurestwo \iffdef \forall s_0, s_1.~ (s_0, s_1) \in P^\times \implies s_0 \in M_{s_0, s_1}
        \\
        &~\begin{aligned}
           \text{where~} M_{s_0, s_1} &\defeq \eventuallyn[][\stepcalc{0}(s_0)][\!
            \setdef*{s_0'}{s_1 \in N_{s_0, s_1, s_0'}}\!] \\
             \text{and}~ N_{s_0, s_1, s_0'} &\defeq \eventuallyn[][\stepcalc{1}(s_1)][\!
              \setdef*{s_1'}{
                (s_0', s_1')\in Q^\times \land ((s_0, s_1), (s_0', s_1')) \in F^\times
                }\!
                ]
              \end{aligned}
    \end{aligned}
  $}
  \end{align*}
\end{definition}

The definitions of $M_{s_0, s_1}$ and $N_{s_0, s_1, s_0'}$ nest $\eventuallynname$ requirements: $M_{s_0, s_1}$ includes all the states where the program $C_0$ reaches a state $s_0'$ within $\stepcalc0(s_0)$ steps, and $N_{s_0, s_1, s_0'}$ includes all the states where the program $C_1$ reaches a state $s_1'$ where $(s_0', s_1')\in Q^\times$ and $((s_0, s_1), (s_0', s_1')) \in F^\times$ hold within $\stepcalc1(s_1)$ steps.

As this definition is based on nested $\eventuallynname$ operators,
thanks to its properties \textsc{conj}, \textsc{comm}, and \textsc{comp},
it follows that the relational Hoare triple $\ensurestwoname$ commutes, is compositional, and allows contract unification.

\begin{lemma}[Commutativity][\makebignumlink{common/relational2.ml}{22}{28}]\label{lemma:commutativity}
  Given precondition $P^\times$, postcondition $Q^\times$, frame condition $F^\times$, and step functions $\stepcalc0, \stepcalc1$, the relational Hoare triple \emph{commutes}:
\[
\ensurestwo \iff \ensurestwo[][\stepcalc1][\stepcalc0][P^S][Q^S][F^S]
\]
where the swapped versions are defined as $X^S \defeq \setdef*{(s_1, s_0)}{(s_0, s_1) \in X^\times}$.
\end{lemma}
This symmetry above ensures that the relational logic is invariant to the program orders, allowing their roles to be interchanged without affecting the triple's validity.


\begin{lemma}[Compositional][\makebignumlink{common/relational2.ml}{393}{415}]\label{lemma:compositional}
  Given three properties $P^\times, R^\times, Q^\times$, two frame conditions $F^\times_0, F^\times_1$, and four step numbers $n_0, n_1, m_0, m_1$, it holds that two relational Hoare triples can be \emph{composed} transitively:
  \begin{align*}
    &\ensurestwo[][\lambda s. n_0][\lambda s. m_0][P^\times][R^\times][F^\times_0] \land
    \ensurestwo[][\lambda s. n_1][\lambda s. m_1][R^\times][Q^\times][F^\times_1] \\
    &\quad \implies \ensurestwo[][\lambda s. n_0 + n_1][\lambda s. m_0 + m_1][P^\times][Q^\times][F^\times_0 \circ F^\times_1]
  \end{align*}
\end{lemma}

Similarly, also the frame condition can be transitively composed.
This is essential in \Section{equivalence} for
the composition of program equivalences.

\begin{lemma}[Compositional of Frame Conditions][\makebignumlink{common/relational2.ml}{515}{562}]\label{lemma:compositional_frame}
%
  Given two preconditions $P, P'$, two postconditions $Q, Q'$, and three frame conditions $F_0, F_1, F_2$, and three step functions $\stepcalc0, \stepcalc1, \stepcalc2$, it holds that two relational Hoare triples can be \emph{composed} transitively with respect to the frame conditions:
      \begin{align*}
    \resizebox{\linewidth}{!}{$
    \begin{aligned}
    \inferrule{
      \ensurestwo[][\stepcalc0][\stepcalc1][P][Q][
        \setdef{
          ((s_0, s_1), (s_0', s_1'))
        }{
          (s_0, s_0') \in F_0 \land (s_1, s_1') \in F_1
        }]
      \\
      \ensurestwo[][\stepcalc1][\stepcalc2][P'][Q'][\setdef{
        ((s_0, s_1), (s_0', s_1'))
      }{
        (s_0, s_0') \in F_1 \land (s_1, s_1') \in F_2
      }]
    }{
      \ensurestwo[][\stepcalc0][\stepcalc2][P \circ P'][Q \circ Q'][\setdef{
        ((s_0, s_1), (s_0', s_1'))
      }{
        (s_0, s_0') \in F_0 \land (s_1, s_1') \in F_2
      }]
    }
    \end{aligned}
  $}
  \end{align*}
\end{lemma}

Lemma~\ref{lemma:compositional_frame} formalizes equivalence transitivity: when a program $C_0$ is equivalent to $C_1$ and $C_1$ is equivalent to $C_2$, then $C_0$ is equivalent to $C_2$.
This Lemma is vital in the equivalence proofs
because proving the correctness of each optimization step independently is easier than directly proving the equivalence of the original and optimized program.


\begin{lemma}[Conjunction][https://google.com]\label{lemma:conjunciton}
  Given two preconditions $P^\times_0, P^\times_1$, two postconditions $Q^\times_0, Q^\times_1$, and a frame condition $F^\times$, two contracts can be unified with a \emph{conjunction}:
\begin{align*}
    &\ensurestwo[][\stepcalc0][\stepcalc1][P^\times_0][Q^\times_0] \land
    \ensurestwo[][\stepcalc0][\stepcalc1][P^\times_1][Q^\times_1]\\
    &\quad
    \implies \ensurestwo[][\stepcalc0][\stepcalc1][P^\times_0\cap P^\times_1][Q^\times_0 \cap Q^\times_1]
\end{align*}
\end{lemma}



All these properties of our Hoare triples enable us to reason about the behavior of two programs, while maintaining the natural properties of a Hoare logic. \Appendix{ensurestwo} presents the additional properties of our program logic $\logic$, including the weakening and strengthening of pre-, post-, and frame conditions.
Implemented in HOL Light, the core of the relational verification amounts to \num{1704} lines of code.

\subsection{Connection with Unary Hoare Triples}

We compare the relational Hoare triple $\ensurestwoname{}$ with the unary counterpart $\ensuresnname$, demonstrating two key transformations: (1) deriving relational Hoare triples from two unary ones, and (2) extracting a unary Hoare triple from a \emph{hybrid} relational one.
These transformations serve a dual purpose.
First, deriving a relational triple from unary ones enables reasoning about the behavior of two programs by analyzing each independently:
\begin{theorem}[][\makebignumlink{common/relational2.ml}{349}{355}]\label{thm:ensuresn-ensurestwo}
  Given two sets of pre-, post-, and frame conditions $P, P', Q, Q'$, $F, F' $, and two step functions $\stepcalc0, \stepcalc1$, it holds that:
  \begin{align*}
    &\ensuresn[][\stepcalc0] \land \ensuresn[][\stepcalc1][P'][Q'][F'] \\
    &\quad \implies
    \ensurestwo[][\stepcalc0][\stepcalc1][P \times P'][Q \times Q'][F \times F']
  \end{align*}
\end{theorem}

Second, extracting a unary triple from a hybrid relational one allows results obtained in the unary logic to be seamlessly promoted to the relational framework.
A hybrid relational triple is a relational triple where the pre-, post-, and frame conditions relate to unary pre-, post-, and frame conditions, respectively.
The goal is to be able to extract a unary Hoare triple from a relational one; hence: ($i$) the relational precondition should always have a satisfying pair $(s_0, s_1)$ when $s_1$ satisfies the unary precondition; ($ii$) if a pair $(s_0, s_1)$ satisfies the relational postcondition, then $s_1$ should satisfy the unary postcondition; and ($iii$) the frame condition should be satisfied for the product relation whenever the second component satisfies the frame condition of the unary relation.

\begin{definition}[Hybrid Relational Ensures][\makebignumlink{common/relational2.ml}{331}{334}]\label{def:hybrid}
  Given the pre-, post-, and frame conditions for the product relation $P^\times, Q^\times$, $F^\times$, and unary pre-, post-, and frame conditions $P, Q $, $F$, and two step functions $\stepcalc0, \stepcalc1 : \state \to \nat$, a \emph{hybrid} relational Hoare triple, written $\hybrid$,
   holds if:
  \begin{align*}
    &\ensurestwo[][\stepcalc0][\stepcalc1][P^\times][Q^\times][F^\times]
    \\ \quad {} \land {} &
    \forall s_1.~ s_1 \in P \implies \exists s_0.~ (s_0, s_1) \in P^\times \tag{$i$}
    \\ \quad {} \land {} &
    \forall s_0, s_1, (s_0, s_1) \in Q^\times \implies s_1 \in Q \tag{$ii$}
    \\ \quad {} \land {} &
    \exists F'.~ \forall s_0, s_1, s_0', s_1'.~ \left(
      \begin{aligned}
        &((s_0', s_1'), (s_0, s_1)) \in F^\times \iff {} \\
        &\quad
          (s_0', s_1') \in F' \land (s_0, s_1) \in F
      \end{aligned}\right) \tag{$iii$}
  \end{align*}
\end{definition}

Employing the hybrid relational triple $\hybridname$ (with the prefix \emph{{\color{seabornRed}h}} denoting ``hybrid'') simplifies the verification process and makes the logic more robust. For instance, it enables translating correctness proofs for one program to another, equivalent program without having to reprove them, saving time and effort.
The next result shows that a hybrid relational Hoare triple can be transformed into a unary Hoare triple.

\begin{theorem}[][\makebignumlink{common/relational2.ml}{329}{346}]\label{thm:hybrid-ensuresn}
  \resizebox{0.819\linewidth}{!}{$
    \hybrid \implies \ensuresn[][\stepcalc1]
  .$}
\end{theorem}

\section{Constant-Time Behavior}
\label{sec:constant-time}

In this section, we show how our relational logic $\logic$ can be applied to reason about constant-time behavior.
As is customary in security analysis, we discriminate between public and private input data by partitioning the state labels into two disjoint sets, i.e., $\labels = \public \cup \private$ and $\public \cap \private = \emptyset$.
Public and private data induce equivalence relations on states, i.e., $\equivpublic$ and $\equivprivate$ respectively.
The public data is accessible to the attacker, while the private data is kept secret.
A program is constant-time if, for the same public input data, any two executions terminate with the same number of clock cycles.
As a result, private data does not influence the execution time of the program.

\paragraph{Constant-Time via Events Accumulation.}
It is not practical to specify constant-time behavior by ensuring that the number of steps is equivalent in executions with the same public data, as it is highly dependent on the underlying hardware.
Due to microarchitectural effects, such as memory access patterns or branch prediction, the number of clock cycles can vary significantly between executions.
Instead, we can safely reason about constant-time behavior by employing a stronger notion of timing security:
a program is \emph{constant-time} if---for the same public input data---any two executions of the program induce {\em the same trace of microarchitectural events}.

To observe these events, we extend the state space $\state$ with an $\texttt{events}$ component in $\eventlabels \defeq \labels \cup \{\texttt{events}\}$. The $\texttt{events}$ component records an ordered list of events, such as memory accesses ($\load~x, n$ or $\store~x, n$; where $x$ is the accessed address and $n$ the operation size in bytes), and branch jumps ($\branch~x, y$; where $x$ and $y$ are the current and the destination program counter, respectively),
Any other variable-time instruction, such as division or floating point operations can also be included in the event trace. In our case studies, these operations are intentionally left unresolved by the operational semantics, and thus not included in the event trace.
These events are public data, i.e., $\texttt{events} \in \public$. The extended state space is $\eventstate$ with operational semantics $\stepsymbolevents$. For instance, loading a memory address $x$ into a 16 bit register $r$ collects a load event of 2 bytes:
  \begin{align*}
    \resizebox{\linewidth}{!}{$
    \begin{aligned}
    \inferrule*[right=Load]{
        s(\instr) = i \quad
        \decode[s][i][] = \texttt{load}~r \texttt{,}~x \\
        s(\memory_x) = v \quad
        s(\texttt{events}) = e \quad
        \length[](r) = 16
    }{s\stepevents s\left[r \mapsto v,~
     \texttt{events} \mapsto (e~\concat~\load~x,2),~
     \instr \mapsto i+\length(\texttt{load}~r \texttt{,} x)\right]}
    \end{aligned}
  $}
  \end{align*}
Therefore, we are now able to specify constant-time behavior by ensuring that the list of microarchitectural events is the same in both executions.
Our approach can be easily extended to include other side-channels, such as power consumption.

While we do not include opcode-level information in our events, instruction opcodes can influence the number of cycles (e.g., the \texttt{cbz} and \texttt{b.ne} instructions in \arm{}).
This relies on an assumption that a program is public information, and therefore the events do not need to carry opcode information.
This assumption can be broken if a program 
runs assembly instructions that are separately stored in a private input buffer. We 
prove that such things do not happen individually.

\begin{definition}[Constant-Time via Event Accumulation]\label{def:constant-time}
    Let $\stepsymbolevents \subseteq \eventstate \times \eventstate$ be an operational semantics that collects the microarchitectural events, $P \subseteq \eventstate$ be a precondition, $Q \subseteq \eventstate$ a postcondition, $F \subseteq \eventstate \times \eventstate$ a frame condition, and $\stepcalc0, \stepcalc1 : \eventstate \to \nat$ two step functions.
    The program $\program(P)$ is \emph{constant-time} with respect to private data $\private$ if it holds that:
    \begin{align*}
        \ensurestwo*{\stepsymbolevents}{{\stepcalc0}}{{\stepcalc1}}{\setdef{(s_0,s_1) \in P \times P}{s_0(\public) = s_1(\public)}}{
        \setdef{(s_0,s_1) \in Q \times Q}{s_0(\texttt{events}) = s_1(\texttt{events})}}{F \times F}
    \end{align*}
\end{definition}

Note that, by constraining the public data to be equal in the precondition, we also require that states share the same event trace before executing the program.

\begin{example}
    The program $\progb{}$ in \Figure[right]{running-example} is constant-time with respect to the microarchitectural events of \Definition{constant-time}.
    Indeed, $\progb$ first branches on the length $n$ of the buffers if $n = 0$ at \Line{compare:cbz}; otherwise, it compares the buffers byte-by-byte.
    Assuming registers of 32 bits, each iteration collects two 4-bytes load events: one for each buffer at Lines~\ref{line:compare:k} and \ref{line:compare:x}.
    Then, it branches to start the next iteration at \Line{compare:loop} until the end of the buffers, no matter what the comparison result is.
    Hence, for any public input value, $\progb$ induces the same event trace.

    In contrast, the program $\proga$ in \Figure[left]{running-example} is not constant-time since the event trace may be different for two executions.
    Consider the following counterexample, where $n = 1$ and the buffers are $k = 10$ and $x = 20$. In memory, the two executions contain $s_0(\memory_{10}) = 0$ and $s_0(\memory_{20}) = 0$; and $s_1(\memory_{10}) = 0$ and $s_1(\memory_{20}) = 1$ respectively.
    The two traces differ at the first mismatch, as the loop in the second execution is terminated early.
    For brevity, the following event traces are simplified omitting the address of branch instructions with the evaluation of the condition:
    \begin{align*}
        \resizebox{\linewidth}{!}{$
        \begin{aligned}
            s_0(\texttt{events}) &= [\branch~\false, \load~10,4, \load~20,4, \branch~\false, \branch~\false] \\
            s_1(\texttt{events}) &= [\branch~\false, \load~10,4, \load~20,4, \branch~\true]
        \end{aligned}
        $}
    \end{align*}
\end{example}

\paragraph{Constant-Time via Unary to Relational Embedding.}
We can employ unary Hoare logic to prove constant-time behavior by showing that private data does not influence the event trace generated during program execution. In other words, it is sufficient to provide a witness trace that depends only on public data.

\begin{definition}[Constant-Time via Unary to Relational Embedding]\label{def:constant-time-1}
    Let $\stepsymbolevents$ be an operational semantics that collects the microarchitectural events, $P$ be a precondition, $Q$ a postcondition, and $F$ a frame condition.
    The program $\program(P)$ is \emph{constant-time} with respect to private data $\private$ if there exists a function $f: \eventstate(\public) \to \events$ such that:
    \begin{align*}
      \forall v_{\publicname}, e_0.~
        \ensuresn*{\stepsymbolevents}{\stepcalc{}}{
            \setdef{s \in P}{
                v_{\publicname} = s(\public) \land e_0 = s(\texttt{events})
            }
        }{
            \setdef{s \in Q}{
                s(\texttt{events}) = e_0 \concat f(v_{\publicname})
            }
        }{
            F
        }
    \end{align*}
    where $\eventstate(\public)$ is the partial projection of states $\eventstate$ on public data $\public$, and $\concat$ is the list concatenation.
\end{definition}

This approach eliminates the need to run the symbolic simulation tactic twice, but requires providing an explicit witness for the event trace function $f$. Since this approach proves a statement about a single program execution, the proof structure is very similar to the correctness proof. Therefore, we can merge the two proofs for correctness and constant-time behavior into a single one; thus eliminating the computational effort of checking each proof separately and greatly reducing the overhead of writing and maintaining them. We can retrieve the relational definition by instantiating \Theorem{ensuresn-ensurestwo} with two instances of the same \ensuresnname\ proof, renamed accordingly.

\begin{example}
    Using list comprehension, for a given public input $v_{\publicname}$,
    the witness $f$ for the program $\progb$ in \Figure{running-example} is defined as:
    \[
        [\branch~(v_{\publicname}(\texttt{n}) = 0)] \concat \listdef*{
            \load~(v_{\publicname}(\texttt{x}) + v_{\publicname}(\texttt{n}) - 1 - i), 4 \\
            \load~(v_{\publicname}(\texttt{y}) + v_{\publicname}(\texttt{n}) - 1 - i), 4 \\
            \branch~(i < v_{\publicname}(\texttt{n}))
        }{
            i \in [0, v_{\publicname}(\texttt{n}))
        }
    \]
    where $\texttt{n}$, $\texttt{x}$, and $\texttt{y}$ are public data and therefore accessible in $v_{\publicname}$.
\end{example}

Note that, routines in the \bignum{} library can be proven constant-time by instantiating either \Definition{constant-time} or \Definition{constant-time-1}, the two are equivalent.

\section{Equivalence Checking}
\label{sec:equivalence}

In this section, we demonstrate the application of our relational Hoare logic framework to equivalence checking between performance and verification-friendly implementations of the same routine in the \bignum{} library.

\paragraph{Equivalence between Two Programs.}

Two programs are considered functionally equivalent if they produce the same output states starting from equivalent input states.
When dealing with assembly-level programs, we must carefully define what it means for two states to be ``equal''.
For instance, two equal input states should not require the exact same code in memory; otherwise, only identical programs could be compared.
Similarly, because the calling convention allows callee-save registers to hold different values, the value of these registers should not be constrained.

On the output side, certain registers or memory regions may differ if they are not designated as outputs.
For example, eliminating dead stores to the stack frame is a valid optimization because the stack frame is not used after function returned.
Two equivalent output states must allow those parts of memory to contain different data.

As a consequence, the equivalence checking takes as a parameter the equivalence relations $\equivin\subseteq\state\times\state$ and $\equivout\subseteq\state\times\state$ that define when input and output states are considered equivalent. This relation has to be defined manually for each pair of programs to be compared.

\begin{example}\label{ex:equivalence}
    Consider the two programs $\proga$ and $\progb$ in \Figure{running-example}.
    Assuming a proof of correctness for $\proga$ already exists, our goal is to prove that the secure constant-time version is functionally equivalent to the original program, without needing to reprove the correctness of $\progb$ from scratch.
    To do so, we define the input equivalence $\equivin$, relating the program counter, input registers, and relevant part of the memory as follows:
    \[
        \equivin = \maychange\left(\labels \setminus \left(
            \begin{aligned}
                &\{\instr,
                    \texttt{n}, \texttt{x}, \texttt{y}
                \} \cup {} \\
                &\setdef{\memory_i}{i \in [x, x+n) \lor i \in [y, y+n)}
            \end{aligned}
        \right)\right)
    \]
    Note the use of the $\maychange$ operator to define a relation that allows two states to differ in all labels but the ones specified.
%
    For output equivalence $\equivout$, we relate only the output register, i.e.,
    $\equivout = \maychange(\labels \setminus \{\texttt{res}\})$.
\end{example}

\begin{definition}[Equivalence]\label{def:equiv}
    Let $P_0, P_1 \subseteq \state$ be two preconditions, $Q_0, Q_1 \subseteq \state$ two postconditions, $F_0, F_1 \subseteq \state \times \state$ two frame conditions, and $\stepcalc0, \stepcalc1 : \state \to \nat$ two step functions.
    Given the input and output equivalences $\equivin, \equivout \subseteq \state \times \state$, the programs $\program(P_0)$ and $\program(P_1)$ are \emph{equivalent} if it holds that:
    \begin{align*}
        \ensurestwo*{}{{\stepcalc0}}{{\stepcalc1}}{
            \setdef{(s_0,s_1) \in P_0 \times P_1}{s_0 \equivin s_1}
        }{
            \setdef{(s_0,s_1) \in Q_0 \times Q_1}{s_0 \equivout s_1}
        }{
            F_0 \times F_1
        }
    \end{align*}
\end{definition}

\begin{example}
    We can prove that the constant-time program $\progb$ is equivalent to the original program in $\proga{}$ by applying \Definition{equiv} with the input and output equivalences defined in \Example{equivalence}.
    Along the lines of the pre- and postconditions defined in \Section{constant-time}, we define:

    \noindent
    \scalebox{0.961}{%
    \parbox{\textwidth}{%
    \begin{gather*}
        P' = \setdef*{s}{s(\texttt{n}) = n \land s(\texttt{x}) = x \land s(\texttt{y}) = y \land {} \\
        \forall i \le n.~ s(\memory_{x+ i}) = \vec{x}_i \land s(\memory_{y+ i}) = \vec{y}_i},\\
            P_0 = \setdef*{s\in P'\!}{\!\alignedto[s][i_0][\progb]}\!,
            P_1 = \setdef*{s\in P'\!}{\!\alignedto[s][i_0][\proga]}\!,\\
            Q_0 =\! \setdef{s\!}{\!\terminatedto[s][\length\!(\progb)]}\!,
            Q_1 =\! \setdef{s\!}{\!\terminatedto[s][\length\!(\proga)]}\!,\\
        F_0 = \maychange(\{\instr, \texttt{n}, \texttt{xn}, \texttt{yn}\}),\\
        F_1 = \maychange(\{\instr, \texttt{n}, \texttt{xn}, \texttt{yn}, \texttt{diff}, \texttt{temp}\}),\\
            \stepcalc0(s) = \largestprefix_n(s, x, y), \text{ and }
        \stepcalc1(s) = s(\texttt{n}),
    \end{gather*}
    }}

    \noindent
    where $\largestprefix_n(s, x, y)$ is the length of the largest prefix among the two given memory addresses $x$ and $y$ of length $n$.
    In conclusion, \Definition{equiv} provides the specification for the equivalence proof between the two programs.
\end{example}

\paragraph{Composition of Program Equivalences.}

We slightly abuse notation and define $\equivensuresname$ as a shorthand for the equivalence of two programs $C_0, C_1$ with $\equivin$ in the precondition starting
from $\pc_0, \pc_1$, eventually reaching $\equivout$ in the postcondition
at $\pc_0', \pc_1'$:
\begin{math}
    \equivensures
\end{math}.
Notably, \Lemma{compositional} proves that the sequential composition of two equivalence proofs is sound if $\forall s, s'.~s \equivout s' \implies s \equivin' s'$. Formally, the \emph{sequential composition} of two equivalences is defined as follows:
\begin{align*}
  \resizebox{\textwidth}{!}{$
    \inferrule{
        \equivensures[][C_0][C_1][\equivin][\equivout][\pc_0][\pc_0'][\pc_1][\pc_1']
        \quad
        \equivensures[][C_0][C_1][\equivin'][\equivout'][\pc_0'][\pc_0''][\pc_1'][\pc_1'']
    }{
        \equivensures[][C_0][C_1][\equivin][\equivout'][\pc_0][\pc_0''][\pc_1][\pc_1'']
    }
    $}
\end{align*}



\Lemma{compositional_frame} instead proves the soundness of the transitive composition of two equivalences, only if the result input and output equivalences preserve the existence of an intermediate state, i.e., $s \equivin' s' \iff \exists s''. (s \equivin s'' \land s'' \equivin' s')$ and $s \equivout' s' \iff \exists s''. (s \equivout s'' \land s'' \equivout' s')$.
Formally, the \emph{transitive composition} of two equivalences is defined as follows:
  \begin{align*}
    \resizebox{\linewidth}{!}{$
    \begin{aligned}
    \inferrule{
        \equivensures[][C_0][C_1][\equivin][\equivout][\pc_0][\pc_0'][\pc_1][\pc_1']
        \quad
        \equivensures[][C_1][C_2][\equivin_1][\equivout_1][\pc_1][\pc_1'][\pc_2][\pc_2']
    }{
        \equivensures[][C_0][C_2][\equivin'][\equivout'][\pc_0][\pc_0'][\pc_2][\pc_2']
    }
\end{aligned}
$}
\end{align*}


\paragraph{Combining Equivalence and Correctness Proofs.}

In the following, we show how to reuse a correctness proof of an original program to obtain a correctness proof of an optimized program through program equivalence.
Indeed, given the functional correctness of the original program in the form of an $\ensuresnname$ proof, we can apply it to the optimized program by proving the equivalence of the two via the relational Hoare triple $\ensurestwoname$.
The correctness proof of the optimized program is given in the form of a hybrid relational Hoare triple $\hybridname$, presented in \Definition{hybrid}.

\begin{theorem}[Transfer of Correctness through Equality]\label{thm:correctness-equality-to-hybrid}
    \begin{align*}
        &\ensuresn[\stepsymbol][\stepcalc0] \land \ensurestwo[\stepsymbol][\stepcalc0][\stepcalc1][P^\times][Q^\times][F^\times]
        \\
        &\implies \hybrid*{\stepsymbol}{\stepcalc0}{\stepcalc1}{
            \setdef{(s_0,s_1)\in P^\times}{ s_0 \in P}
        }{
            \setdef{(s_0,s_1)\in Q^\times}{ s_0 \in Q}
        }{
            \setdef{((s_0,s_1),(s_0',s_1'))\in F^\times}{ (s_0,s_0')\in F}
        }{
            P
        }{
            Q
        }{
            F
        }
    \end{align*}
\end{theorem}


Let $P_0\subseteq\state$ be the precondition, $Q_0\subseteq\state$ the postcondition, $F_0\subseteq\state\times\state$ the frame condition, and $\stepcalc0: \state \to \nat$ the step function.
We state functional correctness as:
\begin{math}
    \ensuresn[][\stepcalc0][
        \setdef{s \in P_0}{s(\pc) = x_0}
    ][
        \setdef{s \in Q_0}{s(\pc) = x_\omega}
    ][
        F_0
    ]
\end{math}.
Afterwards, from \Definition{equiv}, given the two input-output equivalences $\equivin$ and $\equivout$, the equivalence between two programs is achieved by proving:
\begin{align*}
    \ensurestwo*{}{{\stepcalc0}}{{\stepcalc1}}{
        \setdef{(s_0,s_1) \in P_0 \times P_1}{s_0 \equivin s_1}
    }{
        \setdef{(s_0,s_1) \in Q_0 \times Q_1}{s_0 \equivout s_1}
    }{
        F_0 \times F_1
    }
\end{align*}
where $P_1, Q_1, F_1$ are the pre-, post-, and frame conditions of the second program, respectively.
\Theorem{correctness-equality-to-hybrid} transfers the correctness and equivalence proofs to the following hybrid relational Hoare triple:
\begin{align*}
    \resizebox{\textwidth}{!}{$
    \begin{aligned}
    \hybrid*{}{\stepcalc0}{\stepcalc1}{
        \setdef{(s_0,s_1) \in P_0 \times P_1}{s_0 \equivin s_1 \land s_0(\pc) = x_0}
    }{
        \setdef{(s_0,s_1) \in Q_0 \times Q_1}{s_0 \equivout s_1 \land s_0(\pc) = x_\omega}
    }{
        F_0 \times F_1
    }{
        \setdef{s \in P_1}{s(\pc) = x_0}
    }{
        \setdef{s \in Q_1}{s(\pc) = x_\omega}
    }{
        F_1
    }
    \end{aligned}
    $}
\end{align*}
Finally, by applying \Theorem{hybrid-ensuresn}, we obtain the correctness proof of the new program:
\begin{math}
    \ensuresn[][\stepcalc1][
        \setdef{s \in P_1}{s(\pc) = x_0}
    ][
        \setdef{s \in Q_1}{s(\pc) = x_\omega}
    ][
        F_1
    ]
\end{math}.
In \Appendix{equivalence}, we provide the steps required to promote a correctness proof that was originally written via the $\ensuresname$ operator---without an explicit number of steps---to a proof that uses the $\ensuresnname$ operator.
The majority of functional correctness proofs already available in the \bignum{} library are written using the $\ensuresname$ operator.
In total, the core of the equivalence checking proofs is \num{2629} lines of HOL Light code.

\section{Obtaining Proofs for the hol-bignum Library}
\label{sec:evaluation}


\newcommand*\bignumcopy{\texttt{bignum\_copy}}
\newcommand*\bignuminvp{\texttt{bignum\_inv\_p25519}}

\subsection{Case Study: Bignum Copy and Inversion Modulo Routine.}
We apply the constant-time verification to the \bignum{} library, notably on the copy program of large integers, cf. \bignumcopy{}, and the inversion modulo a prime $p = 2^{255} - 19$, cf. \bignuminvp{}.
The following should provide guidance on which proof approach to apply depending on the program size and complexity.

The \bignumcopy{} routine is relatively small, comprising 16 instructions that copy the content of buffer $k$ to the buffer $z$, padding $z$ with zeros if it is bigger than $k$.
Despite its size, \bignumcopy{} has the most complex program flow in the library, making it a good candidate for constant-time verification.
The functional correctness proof is 180 lines.
The constant-time proof, using \Definition{constant-time}, is 276 lines: it does not require an explicit event trace and is fairly easy to prove correct.
On the other hand, the unary constant-time proof using \Definition{constant-time-1} is 245 lines, and requires an explicit event trace.
Although the event trace is small and intuitive, this parameter makes the proof more complex as it requires a nontrivial induction on list comprehensions.
Notably, we can combine correctness and constant-time proofs together via \Theorem{ensuresn-ensurestwo} in a single, 277-line proof, which yields the lowest proof size overhead.

The \bignuminvp{} routine instead is a 1033-instruction program that finds the inverse of a big integer modulo a prime $p = 2^{255} - 19$.
The functional correctness proof is 2303 lines long.
The constant-time proof, using the unary embedding of \Definition{constant-time-1} combining correctness and constant-time proofs, is 2633 lines long.
Most of the additions in the proof are due to the explicit definition of the event trace, which contains 90 memory events alone.
However, after defining the event trace, extending the correctness proof with the constant-time proof was effortless.
All the mechanized proofs are available in the artifact.\footnote{\url{https://doi.org/10.5281/zenodo.15309209}}
In future work, we plan to automate the generation of the event trace, which will significantly reduce the required level of manual effort.

\subsection{Case Study: Elliptic Curves and Montgomery Reduction.}
We utilized program equivalence to verify the functional correctness of optimized implementations for (1) \emph{field and point operations} of NIST elliptic curves
(specifically, curves P-256, P-384, and P-521), and (2) \emph{Montgomery reduction}, an algorithm that allows efficient modular arithmetic when the modulus is large.
These optimizations were achieved using an \emph{autovectorizer}, a constraint solver-based instruction scheduler called SLOTHY \citep{slothy}, and the point operations of NIST curves were optimized using a custom memory instruction optimizer for the \arm{} architecture.
We also have similar equivalence checking tactics for the \x86 architecture.
Overall, we checked the equivalence for 15 pairs of arithmetic routines, amounting to a total of 19k lines of proofs.

The autovectorizer replaces sequences of 64-bit scalar multiplication instructions, such as \texttt{mul} and \texttt{umulh}, with their equivalent NEON vector instructions. This optimization targets the \arm{} Neoverse N1 architecture, whose microarchitecture contains only one multiplication pipeline. The \texttt{mul}/\texttt{umulh} instructions stall this pipeline for a few cycles when executing scalar multiplication instructions.
SLOTHY employs a constraint solver and cost model to find the optimal instruction scheduling, significantly reducing these stalls.
Specifically, SLOTHY improves the scheduling of straight-line code in the main basic blocks of NIST curves' field operations, and also improves the software pipelining optimization in the main loop of the Montgomery reduction.
The memory instruction optimizer performs two key tasks in the \arm{} architecture: store-to-load forwarding and dead store elimination. Store-to-load forwarding replaces load instructions with stored values, eliminating redundant memory accesses. Dead store elimination removes store instructions with results that are never used.

\newcommand*\equivstepstac{\texttt{EQUIV\_STEPS\_TAC}}
\newcommand*\stepsabbrevtac{\texttt{STEPS\_ABBREV\_TAC}}
\newcommand*\stepsrewritetac{\texttt{STEPS\_REWRITE\_TAC}}
\newcommand*\stepsabbrevrewritetac{\texttt{STEPS\_}\{\texttt{ABBREV,REWRITE}\}\texttt{\_TAC}}

\paragraph{Tactics for Program Equivalence Proofs.}
To automate the writing of equivalence proofs, we developed proof tactics that can be used for two different classes of optimizations: small localized updates and instruction reordering.

For local optimizations that update only small portions of the original program, such as autovectorization, we implemented the tactic $\equivstepstac$.
This tactic
takes as input a list
of line ranges and annotations describing whether each range is optimized or left identical.
For the identical portion, the tactic performs lock-step symbolic simulation and eagerly abbreviates the common outputs of the instructions with
fresh variables to avoid exponential explosion of the sizes of the output expressions.
For optimized ranges, the tactic employs stuttering
simulation, which executes the corresponding sections of each program step-by-step.
To help $\equivstepstac$ automatically converge on complex cases, users can register custom bit-vector equality theorems for output expressions.

For optimizations involving instruction reordering, we implemented two additional tactics: $\stepsabbrevtac$ and $\stepsrewritetac$.
The first tactic performs stuttering symbolic simulation for the first program,
storing the symbolic output expressions to an OCaml array.
The second tactic takes as input an instruction index mapping between the two programs, along with the symbolic output generated by $\stepsabbrevtac$.
Then, it simulates the second program step-by-step, proving that the symbolic output of each instruction is equal to the symbolic expression in the first program, according to the instruction mapping.

\paragraph{Software Pipelining of Montgomery Reduction.}

The Montgomery reduction is heavily used in cryptographic operations performing modular exponentations.
Its original implementation in the \bignum{} library includes a nested loop structure, where the outer loop consists of three basic blocks: loop entry, inner loop (which consists of a single basic block), and loop exit.
A faster version was achieved by: caching repetitive calculations, vectorizing \texttt{mul} and \texttt{umulh} in all basic blocks, applying software pipelining optimizations to the inner loop, and rescheduling instructions using SLOTHY.


We verified the functional correctness of the optimized Montgomery reduction by \emph{transitively} composing equivalence proofs with the original correctness proof after each optimization stage.
For each optimization, we applied \emph{sequential} composition of equivalences between each basic block pair and induced the equivalence of the whole loop.
In the case of software pipelining, which transforms the control flow graph by adding loop prologue and epilogue blocks, equivalence between inner loops was proven and composed with equivalences for the loop entry and exit blocks.



Overall, the optimized field operations of NIST curves achieved throughput speedups up to 38\%, and integrating these improvements into point operations alongside memory optimizations resulted in up to 23\% throughput gains. These enhancements demonstrate the substantial impact of the new optimizations on performance of the \bignum{} library.

\section{Conclusion}
\label{sec:conclusion}

This work presents a novel relational Hoare logic framework for verifying realistically modelled machine code, while preserving natural properties expected from Hoare-style reasoning. Fully formalized in HOL Light, the framework is applied in two case studies involving the \bignum{} cryptographic library, a key component of a TLS/SSL implementation.
Our results show that the logic scales to large assembly programs and yields practical value in the verification of cryptographic codebases.

While \citet{mazzucato2024a} have investigated constant-time verification for libraries similar to \bignum{}, their approach relies on abstraction-dependent methods through an untrusted computing base to decompile assembly into C. In contrast, our framework operates directly on the assembly level, ensuring higher reliability of the verification results as it reduces the trusted computing base to the minimal core of the HOL Light theorem prover and to the operational semantics implementations.
As future work, we plan to increase coverage of relational properties on the \bignum{} library and improve proof automation to handle repetitive tasks.
As a natural extension to constant-time proofs, we aim to address speculative execution vulnerabilities \cite{kocher2019,cauligi2022}.

\begin{credits}
\subsubsection{\ackname}
We would like to thank Hanno Becker for his help in improving the Montgomery reduction implementation, and the anonymous reviewers of CAV 2025 for their valuable feedback.
Research reported in this publication was supported by an Amazon Research Award, Fall 2023.
\end{credits}

%
%


\bibliography{s2n-bignum}

\newpage
\appendix

\section{Derivation Rules of \ensuresname{}}\label{app:nonrellogic}

\Figure{nonrellogic} shows the derivation rules of \ensuresname{}, including the precondition weakening (\textsc{pre}), postcondition strengthening (\textsc{post}), frame monotonicity (\textsc{frame}), sequencing (\textsc{seq}), branching (\textsc{branch}), and loop elimination rule (\textsc{loop}).
Regarding \textsc{loop}, we require the invariant $I: \nat \to \wp(\state)$ to be defined for each iteration of the loop, and the postcondition $Q$ to be satisfied after the $k$-th iteration for any $k$ strictly greater than 0.

Note that, given a property of states $P$, we denote by $\overline P$ the complement of $P$ in the state space, i.e., $\overline P = \state \setminus P$.

\begin{figure}[t]
  \centering
  \begin{mathpar}
      \inferrule*[right=pre]{P' \subseteq P \\ \ensures}{\ensures[][P'][Q][F]}
      \and
      \inferrule*[right=post]{Q \subseteq Q' \\ \ensures}{\ensures[][P][Q'][F]}
      \and
      \inferrule*[right=frame]{F' \subseteq F \\ \ensures}{\ensures[][P][Q][F']}
      \and
      \inferrule*[right=seq]{\ensures[][P][R][F] \\ \ensures[][R][Q][F']}{\ensures[][P][Q][F \circ F']}
      \and
      \inferrule*[right=branch]{\ensures[][P \cap B][Q][F] \\ \ensures[][P \cap \overline B][Q][F]}{\ensures}
      \and
      \inferrule*[right=loop]{
        \ensures[][P][I(0)][F]
        \\\\
        \forall i \in \nat.~ i < k \implies \ensures[][I(i)][I(i+1)][F]
        \\\\
        \ensures[][I(k)][Q][F]
        }{\ensures}
  \end{mathpar}
  \caption{Derivation rules of the \ensuresname{} predicate in $\nonrellogic$.}
  \label{fig:nonrellogic}
\end{figure}

\section{Challenges in Extending Unary Logic}
\label{app:challenges}
Defining relational Hoare triples in $\logic$ is nontrivial and requires careful consideration to maintain the natural properties expected from a Hoare logic. Below, we present the challenges that arise when extending unary Hoare logic to a relational one and outline how we address these challenges.

\paragraph{Product Relation.}
A straightforward approach to reasoning about two programs is to define their operational semantics as the product of their individual transitions:
\[ (s_0, s_1) \stepprod (s_0', s_1') \iffdef s_0 \step s_0' \lor s_1 \step s_1' \]

However, this approach is problematic. The operational semantics $\stepprodsymbol$ may advance one program indiscriminately, potentially reaching a state where the postcondition $Q$ no longer holds for the pair, even if both programs satisfy $Q$ individually under the precondition $P$.

\begin{figure}[t]
  \centering
\begin{tikzpicture}[->]
  \node[] (A) at (0, 0) {\((0, 0)\)};
  \node[] (B) at (1, 1) {\((0, 1)\)};
  \node[] (C) at (-1, 1) {\((1, 0)\)};
  \node[] (D) at (0, 2) {\((1, 1)\)};
  \draw (A) -> node[below right] {$\stepprodsymbol$} (B);
  \draw (A) -> node[below left] {$\stepprodsymbol$} (C);
  \draw (B) -> node[above right] {$\stepprodsymbol$} (D);
  \draw (C) -> node[above left] {$\stepprodsymbol$} (D);
\end{tikzpicture}
  \caption{Product relation $\stepsymbol^2$}
  \label{fig:product_relation}
\end{figure}

For example, consider the operational semantics $\stepsymbol = \{(0, 1)\}$ where from state $0$ we can reach state $1$ in a single step. The product relation is $\stepsymbol^2$, depicted in \Figure{product_relation}. The postcondition $Q = \{(0, 1)\}$ should eventually be satisfied by the starting pair of states $(0, 0)$ as the first program already satisfies its postcondition at the initial state $0$, and the second program should eventually reach the state $1$. However, from the initial state $(0, 0)$, one possible transition of the product relation is to the state $(1, 0)$ from which the postcondition $Q$ is not reachable anymore. This behavior prevents the logic from proving intuitive properties.

\paragraph{Lockstep Simulation.}
An alternative approach is a lockstep simulation where both programs advance simultaneously:
\[ (s_0, s_1) \steplock (s_0', s_1') \iffdef s_0 \step s_0' \land s_1 \step s_1' \]

While suitable for programs with identical control flow paths, this approach fails for cases where programs traverse diverging control flows, such as in equivalence checking for optimized and unoptimized implementations.

\paragraph{Nested Eventually Operators.}
Another candidate could be to compose two nested eventually operators to account for the two programs' behavior.
The nested eventually operator would be defined as, for any two states $s_0, s_1$:
\begin{align*}
  (s_0, s_1) \in P \implies s_0 \in \eventually[][
    \setdef*{s_0'}{s_1 \in \eventually[][
      \elimset
    ]}
  ]
\end{align*}

However, this approach is not be able to express two natural properties of Hoare logic. First, it looses symmetry as the nesting of eventually operators imposes ordering, meaning that the resulting logic would not commute between the two programs.
For instance, in the constant-time proofs where the two programs are the same, a postcondition $Q$ could be satisfiable while its inverse $\setdef{(s_1, s_0)}{(s_0, s_1) \in Q}$ may not be, which is unnatural as the two programs are equal and would lead to a significant overhead in the verification framework.
Second, it does not provide a compositional definition. Hence, we would not be able to split the verification of two programs into smaller, independent fragments. Increasing the complexity of the verification process.

\section{Additional Properties of \ensurestwoname}
\label{app:ensurestwo}

As a continuation of the discussion in \Section{relational}, we present additional properties of the \ensurestwoname{} predicate in $\logic$.

As expected from a Hoare logic, the \ensurestwoname{} predicate can be weakened in the precondition, strengthened in the postcondition, and extended in the frame, as shown in the rules \textsc{pre}, \textsc{post}, and \textsc{frame}, respectively.

\begin{mathpar}
  \inferrule*[right=pre]{
    \ensurestwo[][\stepcalc0][\stepcalc1][P] \and P' \subseteq P
  }{
    \ensurestwo[][\stepcalc0][\stepcalc1][P'][Q]
  }\\
  \inferrule*[right=post]{
    \ensurestwo[][\stepcalc0][\stepcalc1][P][Q] \and Q \subseteq Q'
  }{
    \ensurestwo[][\stepcalc0][\stepcalc1][P][Q']
  }
  \\
  \inferrule*[right=frame]{
    \ensurestwo[][\stepcalc0][\stepcalc1][P][Q][F] \and F \subseteq F'
  }{
    \ensurestwo[][\stepcalc0][\stepcalc1][P][Q][F']
  }
\end{mathpar}

Additionally, we can derive a stronger Hoare triple by combining a weaker one with a restriction $f \subseteq \state \times \state$ provided that $f$ contains all and only the pairs of states that are related by the frame condition. Formally:

\begin{mathpar}
  \inferrule{
  \ensurestwo
  \\
    \left(\begin{aligned}
      \forall s_0, s_1, s_0', s_1'.~ ((s_0, s_1), (s_0', s_1')) \in F
      \implies ((s_0, s_1) \in f \iff (s_0', s_1') \in f)
    \end{aligned}\right)
  }{\ensurestwo[][\stepcalc0][\stepcalc1][P \cap f][Q \cap f][F]}
\end{mathpar}

\section{Promotion of Unary Hoare Triples without Steps}
\label{app:equivalence}

If the original correctness proof is written in the $\ensuresname$ form, not $\ensuresnname$,
the proof must be promoted to the $\ensuresname$ form first.
For this reason, we introduce the following $\eventuallynatpcname$ property, stronger than $\eventuallynname$, which allows us to promote a proof made via the $\eventuallyname$ operator to $\eventuallynname$ for a program that halts at a specific program counter.

\begin{definition}[Eventually $n$ at $pc$]\label{def:eventually-n-at-pc}
    Given a number of steps $n\in\nat$, two addresses $x_0, x_\omega\in\nat$ for the initial and final program counters, and a precondition $P\subseteq\state$, we define:
    \begin{align*}
        &\eventuallynatpc \defeq {} \\
        &\quad\setdef*{s\in \state}{
            \forall F\subseteq \state \times \state.~s(\instr) = x_0 \land s\in P \\
            \quad
            \implies
            s \in \eventually[\stepsymbol][\setdef{s'}{s'(\instr) = x_\omega \land (s, s') \in F}] \\
            \qquad
            \implies
            s \in \eventuallyn[\stepsymbol][n][\setdef{s'}{s'(\instr) = x_\omega \land (s, s') \in F}]
        }
    \end{align*}
\end{definition}

Note that this $\eventuallynatpcname$ predicate does not hold for an arbitrary assembly program in general because the program execution may go through the postcondition (which is a part of frame $F$) at $x_\omega$, jump back prior to $x_\omega$ and eventually meet the $F$.
In general, there can be multiple $n$s that meet the postcondition in $F$.

For example, let us assume that $\program{(P)}$ is the single line program adding two operators: $\texttt{add x1, x2, x3}$; we define $x_\omega$ equal to $x_0 + 1$.
It would be tempting to state that $\eventuallynatpcname$ holds when $n=1$ because the program counter arrives at $x_\omega$ after a single step.
However, $n=1$ does not satisfy $\eventuallynatpcname$ because there is a possibility that the next instruction is a decodable instruction that jumps back before the addition.
In this case, any number of steps $n \ge 1$ will arrive at $x_\omega$.

To address this problem, we introduce a \emph{stopper}: a byte sequence that fails to decode.
In \bignum, we use the 4-byte zeros ($\texttt{0x00000000}$) as a stopper sequence.
In $\eventuallynatpcname$, we constrain the program of interest to end with this stopper sequence.
With this stopper sequence appended, we can prove $\eventuallynatpcname$ for a reasonable program and $n$. For the previous example, $n=1$ would be valid.


The next result highlights the promotion of $\ensuresname$ to $\ensuresnname$ given a proof of $\eventuallynatpcname$.

\begin{lemma}[Promotion of $\ensuresname$ to $\ensuresnname$]\label{lemma:ensures-to-ensuresn}
    \begin{align*}
        &\eventuallynatpc \implies \\
        &\quad \forall Q, F.~\ensures[\stepsymbol][P^{pc}][Q^{pc}][F]
        \implies \ensuresn[\stepsymbol][\lambda s.n][P^{pc}][Q^{pc}][F]
    \end{align*}
    where $P^{pc} = \setdef{s\in P}{s(\pc) = x_0}$ and $Q^{pc} = \setdef{s \in Q}{s(\pc) = x_\omega}$.
\end{lemma}

As $\eventuallynatpc$ must end with the stopper sequence, also the precondition $P$ of both $\ensures$ and $\ensuresn$ in the lemma above should handle the stopper sequence as well.
We need to consider take care of both when lifting the existing proofs.

We lift existing proofs of $\ensures$, c.f. the left-hand side of the implication in \Lemma{ensures-to-ensuresn}, that do not mention the stopper sequence by the fact that $\ensures[\stepsymbol][P \land P'][Q][F] \implies \ensures[\stepsymbol][P][Q][F]$ holds,
where $P'$ ensures that the stopper sequence ends the program.


The case of the right-hand side of the implication in \Lemma{ensures-to-ensuresn} seems to be more complicated.
Indeed, we cannot apply the same trick as above because $\ensuresn[\stepsymbol][\lambda s.n][P \land P'][Q][F] \implies \ensuresn[\stepsymbol][\lambda s.n][P][Q][F]$ does not hold.
However, we notice that when combining $\ensuresn$ with a program equivalence proof (defined as $\ensurestwoname$), the stopper sequence does not appear in the final specification.
The reason is that this combination creates a hybrid ensure with its $P,Q,R$ left under an existential quantifier.
Thus, after applying Theorem~\ref{thm:hybrid-ensuresn}, the condition regarding the stopper sequence is removed from the final precondition.

\end{document}